\documentstyle[aas2pp4]{article}

\def\la{\mathrel{\mathpalette\fun <}}
\def\ga{\mathrel{\mathpalette\fun >}}
\def\fun#1#2{\lower3.6pt\vbox{\baselineskip0pt\lineskip.9pt
  \ialign{$\mathsurround=0pt#1\hfil##\hfil$\crcr#2\crcr\sim\crcr}}}

\begin{document}
\tighten

\rightline{FERMILAB--Pub--97/381-A}
\rightline{astro-ph/9711110}
\rightline{submitted to {\it Astrophys. J. (Lett.)}}

\title{No Need for MACHOS in the Halo}
\author{Evalyn I. Gates,$^{1,2}$ Geza Gyuk,$^3$ Gilbert P. Holder,$^2$
and Michael S. Turner$^{2,4,5}$}

\vspace{.2in}
\begin{center}

{\it $^1$Adler Planetarium \& Astronomy Museum, Chicago, IL~~60605}\\

\vspace{0.1in}
{\it $^2$Department of Astronomy \& Astrophysics,
The University of Chicago, Chicago, IL~~60637-1433}\\

\vspace{0.1in}
{\it $^3$S.I.S.S.A., via Beirut 2--4, 34014 Trieste, Italy} \\

\vspace{0.1in}
{\it $^4$Department of Physics,
Enrico Fermi Institute, The University of Chicago, Chicago, IL~~60637-1433}\\

\vspace{0.1in}
{\it $^5$NASA/Fermilab Astrophysics Center,
Fermi National Accelerator Laboratory, Batavia, IL~~60510-0500}\\

\end{center}

\begin{abstract}
A simple interpretation of the more than dozen microlensing
events seen in the direction of the LMC
is a halo population of MACHOs which accounts for about
half of the mass of the Galaxy.  Such an interpretation
is not without its problems, and we show that current microlensing
data can, with some advantage, be explained by dark components
of the disk and spheroid, whose total mass is only about
10\% of the mass of the Galaxy.
\end{abstract}


\section{Introduction}

Microlensing surveys probe the baryonic matter in our Galaxy that exists in
dark compact objects (MACHOs).
These surveys have to date focused on fields toward
the LMC and toward the galactic bulge, although current and future
surveys are probing other lines of sight  including the SMC 
(\cite{smceros,smcmacho}).   The LMC events include
microlensing of LMC stars by disk, halo, and spheroid
lenses, as well as LMC-LMC self lensing.  Contributions to the
microlensing optical depth from known populations in each of these
components has been calculated, and their sum is
significantly less (by about a factor of four) than the current
estimate, $\tau_{\rm LMC} = (2.1^{+1.1}_{-0.7}) \times 10^{-7}$ (\cite{macho2yr}).
Microlensing has apparently revealed
a previously undetected dark population of objects.

On the face of it, the simplest explanation is that the dark halo
has a significant (order 20\% to 80\%) MACHO component.  However,
under this assumption, the inferred lens mass is
around $0.5M_\odot$, which is difficult to
reconcile with searches for white dwarfs and subdwarfs
in the Hubble Deep Field and other surveys (\cite{katie1},\cite{flynn}),
and, at the very least, requires a population of objects with
an unusual initial mass
function.  Other studies indicate that there must be 3 to 10 times
more mass in processed and unprocessed gas than in
MACHOs (\cite{fields}).  There is no evidence for such a large amount
of gas, and the huge mass implied may not be consistent with the
mass budget for the Local Group.

In addition, there is evidence that most of the
baryonic dark matter in the Universe exists in the
form of diffuse hot gas, rather than MACHOs.  In
rich clusters the mass in hot, x-ray emitting gas is about
ten times that in luminous matter in cluster galaxies; this
is consistent with the ratio of the fraction of
critical density in baryons (as determined by big-bang
nucleosynthesis) to that in luminous matter.
Further, if the cluster gas fraction is
taken as a fair sample for matter in the Universe, gas
present when clusters formed accounts for the bulk of
the baryons in the Universe (e.g., \cite{clustergas}).
Further, detailed comparison of the
opacity of the Lyman-$\alpha$ forest with numerical simulations
indicate that the total amount of baryons in gas at redshifts
$z\sim 2 - 4$ accounts for all the baryons.  Finally,
hydrodynamical simulations of structure formation indicate that most
of the baryons remain in diffuse hot gas (\cite{mhdsim}).

While microlensing surveys of the LMC probe the halo only out to a
distance of 50\,kpc, the Galactic halo extends to at least
twice this distance.  If MACHOs represent a significant fraction
of the halo,
they alone eat up most of the baryon-mass budget, leaving little room
for baryonic gas.  In the context of the cold dark
matter paradigm, which provides the only currently viable models
for the evolution of structure in the Universe,
it is difficult -- though not impossible -- to understand half the mass
of the Galaxy being in MACHOs (\cite{gt}, \cite{dgt}).

There are alternatives to halo MACHOs.  Sahu (1994) has
argued that the lenses could be in the LMC itself, though it
seems difficult to achieve the measured optical depth (\cite{difficult}).
Zhao has suggested that the
MACHOs reside in a previously undetected dwarf galaxy located in front
of the LMC (\cite{zhao96}) or in tidal debris in the Magellanic Stream
(\cite{zhao97}).  Evidence
has been presented for (\cite{zaritsky}) and against this hypothesis
(\cite{macho97}).  Finally, the possibility that the lenses are more distant
disk stars that lie along the line-of-sight to the LMC because of warping
and flaring of the Galactic disk has also been suggested (\cite{evans}).

Two, more modest, alternatives involve dark extensions of known galactic
populations, the thick disk (\cite{oldthick}, \cite{thick2}) and the
spheroidal (\cite{oldspheroid}).  Such models are appealing
for several reasons.  The disk and the spheroid
contain significant visible populations, and MACHOS would
be the dark, unseen population.  In fact, dynamical studies
indicate a significant dark population in each.  Finally,
such MACHOs will contribute far less to the total baryon
mass in the Galaxy than MACHOs in an isothermal halo.

In this {\em Letter} we explore the viability of the
hypothesis that all the
microlensing events can be explained with dark extensions of the
disk and spheroid stellar populations.  We show that
models of the Galaxy which are
consistent with the bulge and LMC microlensing data and
which have no MACHOs in the dark halo can be constructed.

\section{Methods}

The methods employed are similar to those of \cite{ggt}.  Galactic
models were constructed by varying the Galactic parameters
relevant to the problem independently over a range
of values consistent with observational data. These models were then
constrained by data on the Galactic rotation curve and by the microlensing
results for the bulge.
Surviving models were binned by $\tau_{\rm LMC}$.

Our models consist of five components:
central bulge, thin disk, thick disk, spheroid and dark halo.  The
baryonic dark matter in MACHOs resides in the thick disk and spheroid;
the thin disk represents the non-lensing component (e.g., gas and bright
stars) of the visible disk; and the halo is comprised solely of
nonlensing dark matter, such as cold dark matter.

The thin disk, which does not contribute to microlensing,
was taken to have an exponential profile in both
$z$ and $R$, with scale lengths $h_z =0.3\,$kpc and $R_d=3.5\,$kpc
and total surface mass density
$25\;M_{\odot}{\rm pc}^{-2}$.  We varied the surface mass
density of the thick-disk component, with the requirement that the total
projected mass density within 1.1\,kpc of the Galactic plane be
in the range
$(35-85)\,M_{\odot}{\rm pc}^{-2}$. The scale height of the thick disk was
allowed to vary between $1\,{\rm kpc}$ and $2.5\,$kpc.

We considered two families of thick disks: exponential disks with
radial scale length $R_d$ in the range $(3-4)\,$kpc, and Mestel disks, with
a surface density  $\propto R^{-1}$.  While a thick disk
population is often assumed to have the same radial density profile as
the observed thin disk, there is no a priori reason to expect this,
especially given the different formation histories expected of the two
populations.

Estimates of the power-law index of the spheroid distribution depend
on the tracer population studied, with an index of $-3$ obtained
from star counts (\cite{starcounts}), and from $-3$ to $-3.5$ from globular
cluster counts (\cite{Frenk1982}).  Different amounts
of dissipation may have occurred in the two populations before
formation. We assume a density profile $\rho(r) \propto {b^3/ (b^3
+ r^3)}$ for the spheroid component, with core radius $b <2\,{\rm kpc}$.

Evidence is mounting that the central region of our Galaxy is
dominated by a bar-like object.  It does not contribute to
microlensing to the LMC, but it is important for both microlensing
toward the Galactic center and for its effect on the inner rotation
curve.  We used the G2 model of \cite{dwek}  for the
bar, and its total mass was constrained to
$(1-4)\times 10^{10} \,M_{\odot}$, with long axis chosen to lie
at an angle of $13.4^{\circ}$ with respect to the line-of-sight to the
Galactic center. The exact details of the bulge model are only important
for calculating the microlensing optical depth toward
the bulge and do not affect our conclusions significantly.

The massive dark halo of the Galaxy was represented by a cored
isothermal sphere of nonlensing objects, with core radius
$(2-12)\,{\rm kpc}$.

The two most important Galactic structure parameters are the rotation
speed at the solar circle, $v_o$, and the solar distance from the
Galactic center, $R_0$.  We incorporated the uncertainties in these
parameters by allowing them to vary independently:
$200\,{\rm km/s} \le v_o \le 240\,{\rm km/s}$ and $7.0\,{\rm
kpc}\le R_0\le 9.0 \,{\rm kpc}.$

All models were then subjected to constraints coming from the observed
rotation curve and from microlensing results toward the Galactic
bulge.  The rotation curve was required to be approximately flat (not
be rising or falling by more than $14\%$) between 4 and 18 kpc from
the Galactic center, and to be in the range 150-307 km/s at 50 kpc.

The optical depth toward the Galactic bulge probes (and constrains)
the mass distribution in the inner Galaxy, which in turn
constrains the other components (\cite{ggt}).
We required viable models to have an
optical depth toward Baade's window ($b=-4^\circ,
l=1^\circ)$ greater than $2.0\times 10^{-6}$, which
is conservative lower bound to
$\tau_{bulge}$ based on data from the OGLE (\cite{ogle}) and MACHO
(\cite{machobulge}) collaborations.

\section{Results}

Figures 1-3 characterize the viable models.  The
range of each parameter is shown as a
function of $\tau_{\rm LMC}$.  Viable Galactic models
were found for $\tau_{\rm LMC}$ as large as $2.5\times 10^{-7}$.
Not surprisingly, higher values of $\tau_{\rm LMC}$
have smaller ranges for all of the parameters,
indicating that these models occupy a small portion of the parameter space.

{}Fig.~1 illustrates that large $\tau_{\rm LMC}$ requires
a combination of large local
rotation speed and small Galactocentric distance:
In order to achieve high optical
depths, a model needs as much lensing material as possible
in the inner galaxy.  Measurements of $v_o$ and $R_0$ are not
independent, however.  The combination of Oort's constants
$A-B=v_o/R_0$ is constrained.  An
analysis by Kerr \& Lynden Bell (1986) found $A-B = 26.4 \pm 1.9$
km/s/kpc.  A more recent analysis of Hipparcos proper motions by
Feast \& Whitelock (1997) finds $A-B= 27.19 \pm 0.87$ km/s/kpc.
With the upper limit $A-B \le 30$ km/s/kpc, no
model with an exponential disk can produce $\tau_{\rm LMC}\ga
1.6\times10^{-7}$, and no model with a Mestel disk can produce
$\tau_{\rm LMC}\ga 2.2\times10^{-7}$.

Overall, the Mestel disk can produce a significantly higher optical
depth toward the LMC.  For a given surface density, the contribution
of a Mestel disk to the inner rotation curve is smaller than that of an
exponential disk.  Thus, models with a Mestel disk can have a higher
disk surface density without exceeding limits on the observed rotation
curve.  Further, the spheroid can be heavier in models with a Mestel
disk, again providing more LMC lenses.

The spheroid mass, total MACHO
mass,{\footnote{For the Mestel disk, this involves a cut-off to the
density distribution in order to keep the mass finite. We chose to
truncate the disk at $15\,{\rm kpc}$. The total disk mass scales
linearly with this truncation radius. } and relative contributions of
the disk and spheroid as a function of optical depth to the LMC are
shown in Figs. 2 and 3.  While $M_{\rm spheroid}$ and $\Sigma$ both
show a fair amount of scatter, the combined mass of the lensing
populations is more tightly constrained.  The
MACHO collaboration notices a similar effect for their halo
models (Alcock et al. 1996).

The required spheroid mass increases with $\tau_{\rm LMC}$,
in some cases becoming comparable to the visible mass of the disk.
Estimates of the dynamical mass in the spheroid are roughly in the
range of $(5-7) \times 10^{10} M_{\odot}$ (\cite {dynmass}),
while the luminous mass is considerably less, around $(1-3)
\times 10^{10} M_{\odot}$ (\cite {lummass}).  The
spheroid mass in Mestel disk models is consistent with these
estimates; in exponential disk models these estimates
restrict $\tau_{\rm LMC}
\la 2\times10^{-7}$, as they require a heavier spheroid.

The growing importance of the spheroid for large values of
$\tau_{\rm LMC}$ is also seen in the last panel in Figs. 2 and 3.
Basically there is an upper bound to the amount of lensing that can be
done by the disk, especially an exponential disk, so that a large
optical depth can only come with a substantial spheroid contribution.

It has been suggested that alternate lines of sight can help break the
degeneracy between Galactic models. However,
the high-latitude bulge fields and the SMC do not offer much hope in
this respect. For example, the signature of flattening suggested by
Sackett \& Gould (1993), where the SMC optical depth is enhanced
relative to the LMC for a flattened halo, can also be reproduced by
models with a thick disk, while a very heavy spheroid can produce the
signal expected of a spherical halo. The high-latitude bulge fields
suffer from the problem that there is a fixed amount of optical depth
toward the bulge itself which must be accounted for. We find that any
model that can produce this bulge optical depth also produces roughly
the same optical depth toward high-latitude bulge fields. This is true
for scenarios both with and without MACHOs in the Galactic halo.
Globular clusters hold more promise (\cite{globs}), but their
feasibility as a target for microlensing surveys is still under investigation.


\section{Discussion}

We have found viable Galactic models in which MACHOs in dark
extensions of the thick disk and spheroid alone produce an
optical depth for microlensing toward the LMC of around $2\times
10^{-7}$.  Thus, at present, microlensing does not require
a significant halo MACHO population.

The mass in MACHOs in our Galactic models is only a small fraction of
the total Galactic mass, of order $10\%$.  This is in line
with the evidence that most of the dark baryons are in gaseous
form, and the cold dark matter paradigm, which holds that most of the
mass of the Galaxy should be cold dark matter particles.
On the other hand, in Galactic models where the MACHOs
are a significant fraction of an isothermal halo which extends to 100
kpc or more, MACHOs comprise around almost half of the total mass in
the Galaxy; this is difficult to reconcile with most of the dark matter
being nonbaryonic and most of the baryons being gaseous.

The estimate for the average mass of the lenses in our models is
only slightly lower than that determined for halo MACHO models,
$\langle m \rangle \sim 0.3 M_{\odot}$ ($0.2
M_{\odot}$ for disk lenses and $0.36 M_{\odot}$ for spheroid lenses),
compared with $0.5 M_{\odot}$ for halo lenses.  The puzzle
of what the lenses are remains.  However, it should be noted that
the baryon mass budget problem is far less severe, and the constraints
from direct searches for lenses may be less severe because the
lenses are not distributed like the halo  (\cite{fatdisk}).

Better measurements of $v_0$, $R_0$, $\tau_{\rm LMC}$,
and the Galactic rotation curve
hold leverage in testing the spheroid/heavy disk hypothesis.
Parallax measurements of the lensing events,
which allow an estimate of the distance to the lens, and/or future lensing
surveys towards globular clusters, which probe additional lines of
sight through the Galaxy can also distinguish between
halo and nonhalo models.

\acknowledgments

This work was supported in part by the DOE (at Chicago and Fermilab)
and by the NASA (through grant NAG5-2788).

\section*{Figure Captions}

1.  Range of $v_o$, $R_0$ and the Oort constants ($A-B$)
as a function of $\tau_{\rm LMC}$ for viable  models
with MACHOs in a thick exponential disk (solid line)
and in a thick Mestel disk (dashed line); scale height is 1.5 kpc.

2.  Range of parameters for  viable  models with MACHOs
in a thick exponential disk with scale heights 1.0\,kpc
(solid square), 1.5\,kpc (open square), 2.0\,kpc (solid triangle),
and 2.5\,kpc (open triangle).
>From top to bottom: Spheroid mass; total mass in baryons;
and ratio of the spheroid to disk contributions to $\tau_{\rm LMC}$.

3.  Same as Fig. 2 for thick Mestel disk models.

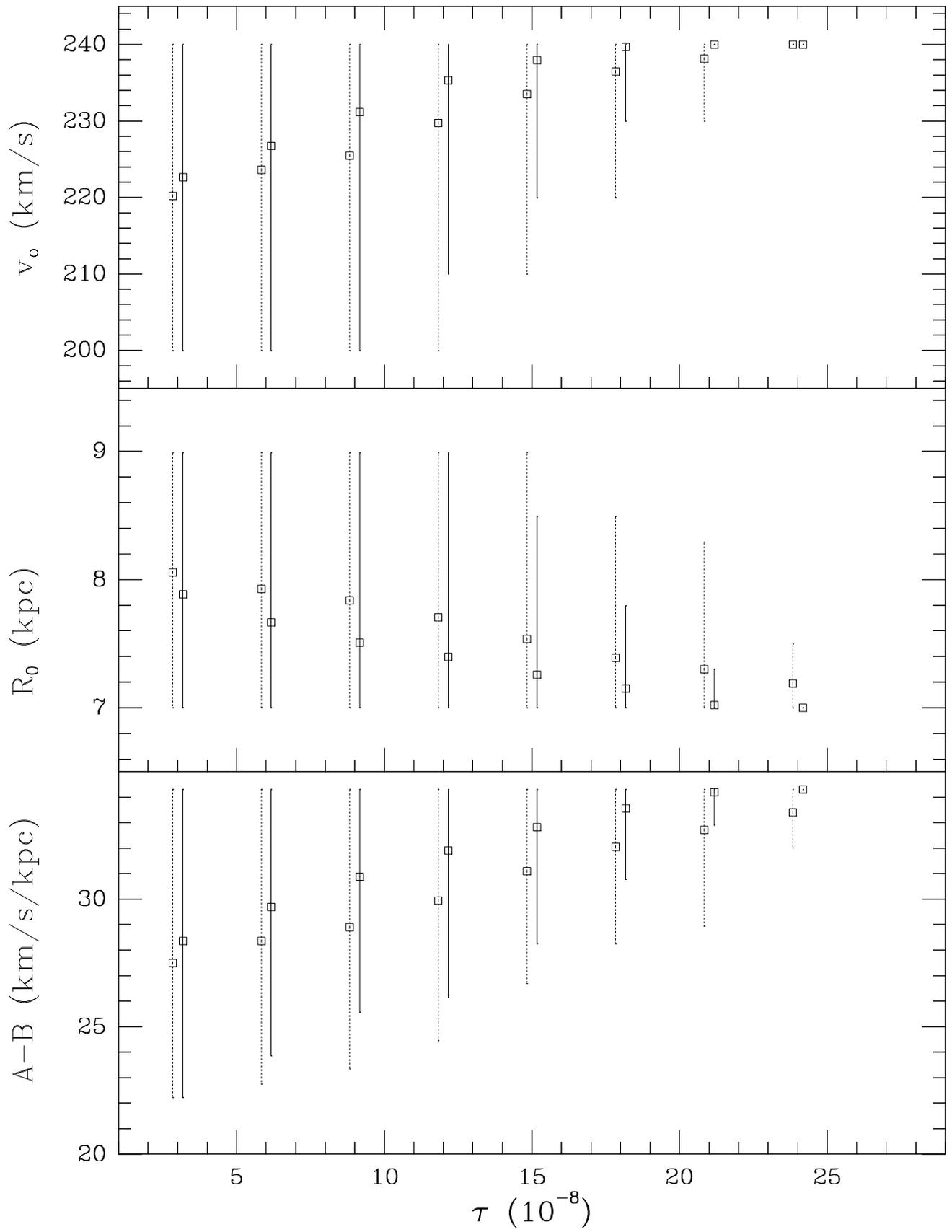
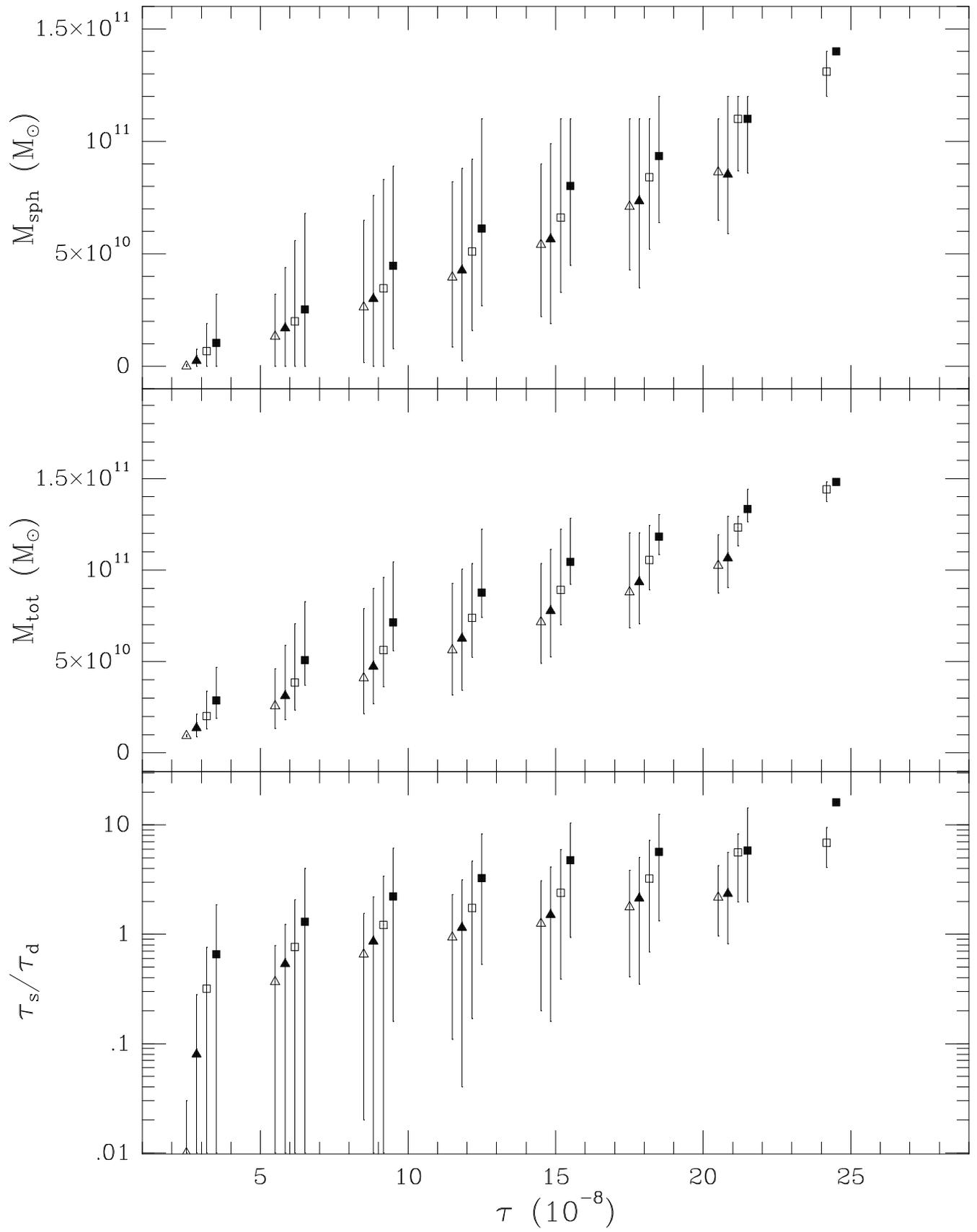
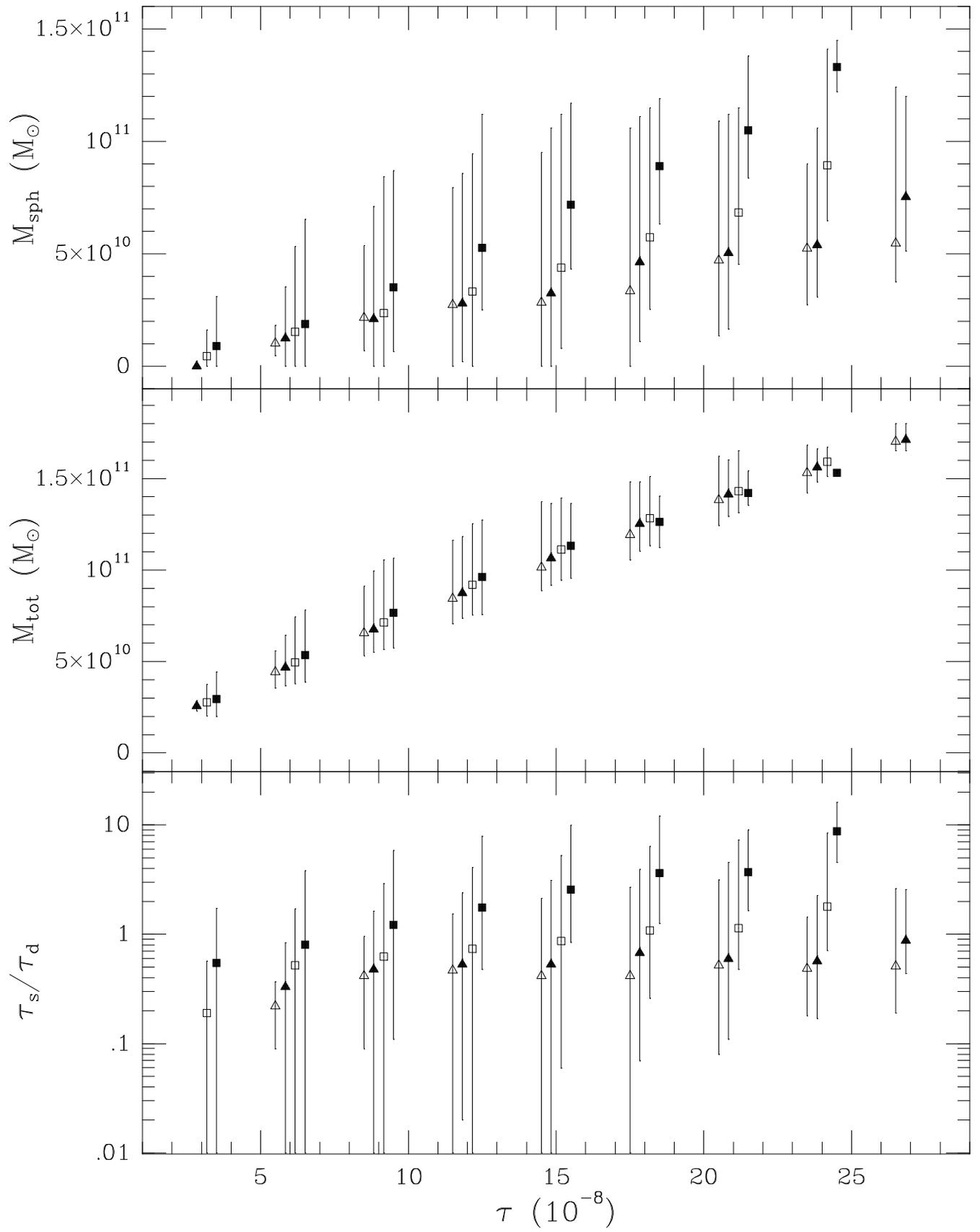
\end{document}